\documentstyle[prb,aps]{revtex}
\def\f{$f$}
\def\k{{\bf k}}
\def\r{{\bf r}}
\def\O{{\cal O}}
\def\H{{\cal H}}
\newcommand{\deff}{\stackrel{{\rm def}}{=}}
\begin{document}
\date{\today}

\title{Theory of Strongly Correlated Electron Systems. \\
  III. Including 
       Correlation Effects into Electronic Structure Calculations}

\author{U. Lundin$^1$, I. Sandalov$^{1,2}$, O. Eriksson$^1$}
\address{$^1$Condensed Matter Theory group, Uppsala University, Box 530,
SE-751 21 Uppsala, Sweden\\
$^2$Dept.\ of Physics, Link\"oping University,
SE-581 83 Link\"{o}ping, Sweden}

\maketitle

\begin{abstract}
In previous papers~\cite{lundin_two_f,lundinsqrtp} we showed that a division 
of the 
\f-shell into two subsystems gives a better understanding of cohesive 
properties as well the general behavior of lanthanide systems. 
In this paper we present 
numerical computations using the suggested method. We show 
that the picture is consistent with most experimental data, e.g.\ the 
equilibrium volume
and electronic structure in general. 
Compared to standard energy band calculations, or to calculations
based on the self interaction correction and LDA+U, in the present approach 
$f$-(non-$f$)-mixing interaction is decreased by spectral weights of the 
many-body states of the $f$-ion are involved. 
\end{abstract}

\section{Introduction}
Strongly correlated electrons show a rich variety of physical properties, 
e.g.\ heavy Fermion
behavior, partial localization and unconventional superconductivity. 
Normal {\em ab initio} electronic structure 
calculations performed for these systems have proven to be unable  
to capture the essential features of these systems. The reason for this is 
mainly 
that the LDA underestimates the strong correlations caused by the local part
of the Coulomb interaction among the electrons. The standard model 
for the lanthanides corrects for this via imposing 
a condition on the \f-electrons which is not present in the standard 
LDA DFT scheme: i) $f$-electrons should be treated as atomic like and
localized; ii) the $f$-shell contains an integral number, $n$, of electrons.
However, it has been argued~\cite{lundinsqrtp}, 
that already the cohesive properties signal a deviation from the 
standard model. 

In a previous paper~\cite{lundinsqrtp}, we presented an idea 
how normal band structure 
calculations could be extended to treat systems with strongly 
correlated electrons, that is an improvement on normal methods such as 
LDA+U, SIC or
treating the f-electrons as atomic like. Indeed it was demonstrated that the 
cohesive properties,
such as e.g.\ equilibrium volume is better described with this method. 
The calculation~\cite{lundinsqrtp} has been performed within the 
linear muffin-tin orbital method within the spherical approximation 
(LMTO-ASA). The comparison with the results of normal LMTO-ASA 
calculation has shown that the latter contains two type of errors: one comes 
from the spherical approximation and another from the underestimated electron 
correlations. Here we have implemented
these many-body corrections in a full-potential 
all-electron relativistic energy band method, in order 
to obtain a description as accurate as possible, and we have 
tested our theory for the light lanthanides. 
These elements are an excellent testing ground for any theory that tries
to incorporate electron correlations, since so much is known experimentally
about them. For this reason, we have chosen to test our theory on these
materials, before addressing more involved systems such as the heavy
Fermion materials, UPt$_3$ and CeCu$_2$Si$_2$ and possibly the
$\delta$-phase of Pu. 

The paper is organized as follows. In Section II we derive the equations
that form the ground for the calculations, in Section III we describe the 
details of the 
numerical implementation, and Section IV contains the results. Finally we 
conclude in 
Section V.

\section{Derivation of the Kohn-Sham like equations}

The main motivations of this work are to derive expressions for corrections
from strong electron correlations to
standard methods of {\em ab initio} methods.
We also want to make an attempt to understand, 
from a many-body theory point of view, the
mechanism of localization of f-electrons, that may be described
phenomenologically by self-interaction correction (SIC),  
and the so called LDA+U method. 
As we shall see the method allows us to capture 
features of correlated systems such as e.g. the correlation narrowing of 
bands. 
The full description of our
approach, as well as notations, will be given in 
ref.~\cite{sandalov_large}, and we present here only some of the
theoretical details, how we have implemented the theory numerically as
well as the results of some of our calculations.

We formulate the problem following the ideology presented 
in ref.~\cite{igor_totale}, namely, 
that the one-to-one correspondence between the ARF (the approximation of 
renormalized Fermions~\cite{igor_totale}) series and the one from 
weak-coupling perturbation 
theory allows to use the same analytical form for the exchange-correlation 
potential. Moreover, our theory includes 
renormalized mixing, hopping, overlap 
matrixes and charge density, by the spectral weights 
of the many-electron ion states that we involve. 
The ARF can be viewed as the Hubbard-I 
approximation which is extended to the case of full Coulomb interaction.
This approximation is the simplest possible, but since it is exact in the 
atomic limit it is expected to give reasonable results in the case of 
very strong correlations.

Now let us establish a connection between the secular equations solved in 
electronic structure calculations (ESC) and Green function methods for 
strongly correlated electrons.
In order to do this  we need to define the Hamiltonians 
for the two problems. First, when solving the Kohn-Sham equation 
we have to solve a secular equation, which
looks like,  
\begin{equation}
\label{eoh0:eq}
(\epsilon \O_{jL,j_1L_1}-h^c_{jL,j_1L_1})\chi_{j_1L_1}={\bf 0},
\end{equation}
where $\O_{jL,j_1L_1}$ is the overlap matrix between the cite-centered 
basis-functions, $\chi_{j_1L_1}$; $j$ is the site index and $L$ is a 
composite index $L=(l,m_l,\sigma)$. $\epsilon_{jL}$ is the energy for the 
orbital, 
and $h^c_{jL,j_1L_1}$ is the Hamiltonian matrix element which will be defined 
below. Let us first consider the \f-electrons as localized.
We, therefore write the 
Hamiltonian for the band structure method with the \f-electrons separated
from the other conduction electrons.
Denoting the operators for \f- and conduction-electrons with $f$ and $c$,
respectively, the ESC Hamiltonian is 
\begin{eqnarray}
\label{bsc_hamiltonian:eq}
\hspace{-0.3cm}&&\H=\sum_{\k,L_1,L_2}h^c_{L_1,L_2}(\k)c^{\dag}_{L_1\k}
          c^{\vphantom{\dag}}_{L_2\k} + 
     \sum_{j_1\nu_1} \epsilon^0 f^{\dag}_{j_1\nu_1}
          f^{\vphantom{\dag}}_{j_1\nu_1}  \nonumber \\ 
&& + \sum_{j_1\nu_1,j_2\nu_2}t^{\nu_1\nu_2}_{j_1j_2}f^{\dag}_{j_1\nu_1}
          f^{\vphantom{\dag}}_{j_2\nu_2} \nonumber \\
&& + \sum_{\k,L_1,j_2\nu_2}V_{j_1L_1,j_2\nu_2}[c^{\dag}_{L_1\k}
          f^{\vphantom{\dag}}_{j_2\nu_2} +     f^{\dag}_{j_2\nu_2}
          c^{\vphantom{\dag}}_{L_1\k}     ],
\end{eqnarray}
where the matrix elements comes from the expectation value of the Hamiltonian 
with respect to the basis functions
\begin{equation}
\label{matrix_elements:eq}
h^c_{L_1,L_2}(\k)=\langle\chi_{L_1}(\k)|{\cal H}|\chi_{L_2}(\k) \rangle.
\end{equation}
The index $L$ is used solely for conduction electrons. 
We denote the \f-\f~ hopping by $t$ and the mixing between $c$ and $f$ by $V$. 
Combined indices for the \f's are denoted with $\nu$.
The operators fulfill the commutation relation
\begin{equation}
\label{comm:eq}
\{a^{\vphantom{\dag}}_{j_1L_1},a^{\dag}_{j_2L_2}\}=
 {\cal O}^{-1}_{j_1L_1,J_2L_2},
\end{equation}
where $a$ is either a $c-$ or an \f-operator. 
The Hamiltonian, Eq.(\ref{matrix_elements:eq}), is a DFT Hamiltonian using 
either LDA or any other approximation for the exchange-correlation potential, 
i.e.\
\begin{equation}
\label{ham_LDA:eq}
{\cal H}=\frac{p^2}{2m}+\int d\r d\r' \frac{\rho(\r)\rho(\r')}{|\r-\r'|}+
         V^{{\rm XC}}(\rho(\r))+V_{en},
\end{equation}
where $V^{{\rm XC}}(\rho(\r))$ is the exchange correlation potential
and $V_{en}$ is the interaction of electrons with nuclei.
From these definitions we can define Green functions which we 
can use to make a comparison to the case for strongly correlated electrons. 

Obviously, the same problem can be written in terms of Green functions (GFs).
Denoting the Fermionic Green functions F and separating 
the $c-$ and \f-electrons we can write the GF, $F$, in super matrix form, 
\begin{eqnarray}
\hspace{-0.3cm}&&\left(
\begin{array}{cc}
F_{L_1L_2}^{cc^{\dagger }}(\k,i\omega ) & 
F_{L_1\nu_2}^{cf^{\dagger }}(\k,i\omega ) \\
F_{\nu_1L_2}^{fc^{\dagger }}(\k,i\omega ) & 
F_{\nu_1\nu_2}^{ff^{\dagger }}(\k,i\omega )
\end{array}
\right) \nonumber \\
&&\deff \frac{1}{i}\left(
\begin{array}{cc}
\langle {\cal T} c^{\vphantom{\dag}}_{j_1L_1}c^{\dag}_{j_2L_2} \rangle &
\langle {\cal T} c^{\vphantom{\dag}}_{j_1L_1}f^{\dag}_{j_2L_2} \rangle \\
\langle {\cal T} f^{\vphantom{\dag}}_{j_1L_1}c^{\dag}_{j_2L_2} \rangle & 
\langle {\cal T} f^{\vphantom{\dag}}_{j_1L_1}f^{\dag}_{j_2L_2} \rangle 
\end{array}
\right)_{\k,\omega},
\end{eqnarray}
where ${\cal T}$ is the chronological ordering operator. Further, 
we denote the time Fourier transform by a small $\omega$ and 
$\k$-space Fourier transform with a small $\k$. Taking into account the 
commutation relations, Eq.(\ref{comm:eq}) in this representation,  
we obtain the equation of motion for the Green functions 
(the $\k$ in ${\cal O}(\k)$ is omitted for convenience):
\begin{eqnarray}
\label{bsc:eq}
&&\left(
\begin{array}{cc}
\lbrack \delta _{LL_2}i\omega -{\cal O}_{L,L_1}^{-1}h_{L_1L_2}^c-{\cal O}_{L\nu
_1}^{-1}V_{\nu _1L_2}^{*}] & -[{\cal O}_{L,\nu _2}^{-1}V_{\nu _2\nu
_1}+{\cal O}_{LL_1}^{-1}V_{L_1\nu _2}] \\
\lbrack {\cal O}_{\nu L_1}^{-1}h_{L_1L_2}^c+{\cal O}_{\nu \nu _1}^{-1}
V_{\nu _1L_2}] &
[\delta _{\nu \nu _1}i\omega -{\cal O}_{\nu \nu _1}^{-1}t_{\nu _1\nu _2}-
{\cal O}_{\nu L_1}^{-1}V_{L_1\nu _2}]
\end{array}
\right)
\left(
\begin{array}{cc}
F_{L_2L^{\prime }}^{cc^{\dagger }}(\k,i\omega ) & F_{L_2\nu ^{\prime
}}^{cf^{\dagger }}(\k,i\omega ) \\
F_{\nu _2L^{\prime }}^{fc^{\dagger }}(\k,i\omega ) & F_{\nu _2\nu ^{\prime
}}^{ff^{\dagger }}(\k,i\omega )
\end{array}
\right) \nonumber \\
&&=i\left(
\begin{array}{cc}
{\cal O}_{L,L^{\prime }}^{-1}   & {\cal O}_{L\nu ^{\prime }}^{-1} \\
{\cal O}_{\nu L^{\prime }}^{-1} & {\cal O}_{\nu \nu ^{\prime }}^{-1}
\end{array}
\right).
\end{eqnarray}
Thus, if we multiply both sides of Eq.(\ref{bsc:eq}) by the overlap matrix, 
then, the inverse of the GF coincides 
with the bracket in Eq.(\ref{eoh0:eq}), and the corresponding 
secular determinant gives a band structure
coinciding with Eq.(\ref{eoh0:eq}). 
Let us now move to the problem of strongly correlated electrons. 
We will work with the multiple-orbital
periodical Anderson model (MO-PAM) in terms of Hubbard operators. 
The latter are defined as projection (transition) operators acting on the 
many-body states of an ion, $i$, in
the position $\bf{R}_i$, as,
\begin{equation}
X^{\Gamma,\Gamma'}_i\deff|i,\Gamma\rangle\langle \Gamma',i|.
\end{equation}
Here $|\Gamma_n\rangle$ is an n-particle state constructed out of orbitals 
from 
the ESC, $|\Gamma_n\rangle=f^{\dag}_1f^{\dag}_2\ldots f^{\dag}_n|0\rangle$. 
To simplify the notations we denote the Fermi-like transitions  
$[\Gamma,\Gamma']$ with Roman letters $a,b,\ldots$. 
The commutator relation for the $X$-operators are written in terms of
structure constants $\epsilon$ of the algebra
\begin{equation}
\{X^a,X^{\bar{b}}\}=\sum_\xi \epsilon_\xi^{a,\bar{b}}Z^\xi,
\end{equation}
where $Z^{\xi}$, denotes a Bose-like transition, changing the particle number 
with an even number. 
The Fermion $f$-operators can be expressed in terms of $X$-operators (or
vice versa), i.e.,
\begin{equation}
\label{Xfexpansion:eq}
f_{j\nu} = \sum_{\Gamma,\Gamma'} \langle\Gamma|f_{j\nu}|\Gamma'\rangle
X_j^{\Gamma',\Gamma} = \sum_{a}(f_{j\nu})^aX_j^a.
\end{equation}

The model for the strongly correlated electrons, 
the multiple-orbital periodical Anderson model (MO-PAM) 
(Eq.(\ref{bsc_hamiltonian:eq}) with hopping $t$=0), is expressed in 
terms of $X$-operators using the relation in Eq.(\ref{Xfexpansion:eq}),  
\begin{eqnarray}
\hspace{-0.3cm}&& {\cal H}=\sum_{\k,L_1}\epsilon_{\k L_1}c^{\dag}_{\k L_1}
                                       c^{\vphantom{\dag}}_{\k L_1} +
         \sum_{ja}E^0_aZ_j^a \nonumber \\
&& + \sum_{\k,L_1,j\nu_2,a} \left [
                        (f^{\vphantom{\dag}}_{j\nu_2})^a 
                             c^{\dag}_{\k L_1}X^a_j + c.c \right ], 
\end{eqnarray}
where the band-structure for the $c$ electrons is $\epsilon_{\k L_1}$ 
and $E^0_a=\epsilon^0(f^{\dag}_{j\nu_2})^a(f^{\vphantom{\dag}}_{j\nu_2})^a$. 
Within the ARF we are dealing with the same effective Hamiltonian as 
Eq.(\ref{bsc_hamiltonian:eq}) 
with the exception that instead of the term $\sim\epsilon^0 f^{\dagger}f$ 
we have to write the term describing the many-electron states of an ion, 
\begin{equation}
{\cal H}^0_f = \sum_{j,\Gamma} E_{j,\Gamma} h_{j}^{\Gamma},
\end{equation}
and $f$-operators should be rewritten in terms of Hubbard operators 
(see ref~\cite{igor_totale}). 
Then the effective Hamiltonian acquires the form of the
multi-orbital periodical Anderson model (MO-PAM).
The equations for the Green functions for the multiple-orbital
periodical Anderson model (MO-PAM) in the Hubbard-I approximation 
within a non-orthogonal basis set have the following form,
\begin{eqnarray} 
&&\hspace*{-0.5cm}\left[ \delta _{jj_2}\delta _{LL_2}i\partial
_t-{\cal O}_{jL,j_1L_1}^{-1}h_{j_1L_1,j_2L_2}^c\right]G_{j_2L_2,j^{\prime 
}}^{c\eta
^{\dagger }}(t,t^{\prime }) \nonumber \\
&&\hspace*{-0.4cm}=i\delta (t-t^{\prime })\langle \{c_{jL},\eta
_{j^{\prime }}^{\dagger }\}\rangle  \nonumber \\
&&\hspace*{-0.4cm}+\left[{\cal O}_{jL,j_2L_2}^{-1}(f_\mu)^a 
E_\Gamma \epsilon _b^{a\Gamma
}+{\cal O}_{jL,j_1L_1}^{-1}V_{j_1L_1,j_2\mu }(f_\mu)^b \right]\nonumber \\
&&\hspace*{-0.4cm}\times G_{j_2b,j^{\prime }}^{X\eta
^{\dagger }}(t,t^{\prime }) \nonumber \\
&&\hspace*{-0.4cm}+{\cal O}_{jL,j_1\mu _1}^{-1}(f_{\mu_1})^a 
V_{j_1\mu ,j_2L_2}^{*}(f_\mu 
^{\dagger
})^{\bar{a}}\epsilon _\xi ^{b\bar{a}}\langle Z_{j_1}^\xi \rangle
G_{j_2L_2,j^{\prime }}^{c\eta ^{\dagger }}(t,t^{\prime }), 
\label{system1:eq}
\end{eqnarray}
\begin{eqnarray}
&&\hspace{-0.5cm}[ i\partial _t-\Delta _a]G_{ja,,j^{\prime }}^{X\eta ^{\dagger
}}=i\delta (t-t^{\prime })\langle \{X_j^a,\eta _{j^{\prime }}^{\dagger
}\}\rangle  \nonumber \\
&&\hspace*{-0.4cm}+\left[{\cal O}_{j\mu ,j_1L_1}^{-1}h_{j_1L_1,j_2L_2}^c+
{\cal O}_{\mu \mu _1}^{-1}
V_{j\mu
_1,j_2L_2}^{*}\right] \nonumber \\
&&\hspace*{-0.4cm}\times(f_\mu ^{\dagger })^{\bar{b}}\epsilon _\xi ^{\bar{b}
a}\langle Z_j^\xi \rangle G_{j_2L_2,j^{\prime }}^{c\eta ^{\dagger
}}(t,t^{\prime })  \nonumber \\
&&\hspace*{-0.4cm}+{\cal O}_{j\mu _1,j_1L_1}^{-1}V_{jL_1,j_2\mu }
(f_{\mu _1}^{\dagger })^{\bar{b}
}\epsilon _\xi ^{\bar{b}a}\langle Z_j^\xi \rangle (f_\mu)^a
G_{j_2a,j^{\prime }}^{X\eta ^{\dagger }}(t,t^{\prime }),
\label{system2:eq}  
\end{eqnarray}
where $\eta$ denotes either a $c$ or a $X$-operator, $\Delta_a$ is the energy 
of the transition $a$ (e.g.\ if $a=\Gamma_{n-1},\Gamma_{n}$ then 
$\Delta_a=E_{\Gamma_n}-E_{\Gamma_{n-1}}$). 
We simplify the model further neglecting the energy differences 
between different transitions in the lower Hubbard sub-band, i.e.,
between $n$- and ($n-1$)-electron states. 
Then all these transitions have the same energy, $\Delta_1$.
The same will be assumed about the upper
transitions, between ($n+1$)- and $n$-electron states; they then 
have the energy 
$\Delta_2.$ We assume that the lower transition lies so much
lower in energy than the bottom of all conduction bands that the overlap
integrals as well as mixing interaction of conduction electrons with these
transitions are equal to zero. Therefore, for 
$\alpha =[\Gamma _{n-1,}\Gamma _n]$, the GFs $G_{j\alpha ,j^{\prime
}}^{Xc^{\dagger }}=0$ and
\begin{equation}
G_{j\alpha ,j^{\prime }\alpha ^{\prime }}^{XX}=\delta _{\alpha
\alpha ^{\prime }}\delta _{jj^{\prime }}\frac{P^\alpha }{i\omega -\Delta
_\alpha }.
\end{equation}
This GF gives the equation for the population numbers
\begin{equation}
N_{\Gamma _n}=(N_{\Gamma _{n-1}}+N_{\Gamma _n})f(\Delta _\alpha ),
\label{poop:eq}
\end{equation}
where $f(\Delta _\alpha )$ is Fermi function for the transition 
$\Delta_\alpha $. Since $\beta \Delta _\alpha \gg 1$, 
$N_{\Gamma _{n-1}}=\exp{(-\beta\Delta_\alpha)}N_{\Gamma_n}$. The 
lower population numbers are negligible and 
only $N_{\Gamma _n}$ and $N_{\Gamma _{n+1}}$ have
nonzero values, while Eq.(\ref{poop:eq}) becomes an identity.
 This allows us to formulate the system of equations for 
GFs which contains only upper
transitions. After Fourier transformations $t-t^{\prime }\rightarrow
i\omega $ and ${\bf R}_j-{\bf R}_{j^{\prime }}\rightarrow \k$, 
the system of equations acquires the following form:
\begin{eqnarray}
&&\left[ \delta _{LL_2}i\omega
-{\cal O}_{L,L_1}^{-1}(\k)h_{L_1L_2}^c(\k)\right]G_{L_2}^{c\eta ^{\dagger }}
(\k,i\omega
)=i\langle \{c_{jL},\eta _{j^{\prime }}^{\dagger }\}\rangle_\k  \nonumber \\
&&+[{\cal O}_{L,\mu }^{-1}(\k)(f_\mu)^aE_\Gamma \epsilon _b^{a\Gamma
}+{\cal O}_{LL_1}^{-1}(\k)V_{L_1\mu }(\k)(f_\mu)^b]G_b^{X\eta ^{\dagger }}
(\k,i\omega )
\nonumber \\
&&+{\cal O}_{L\mu _1}^{-1}(\k)(f_{\mu_1})^aV_{\mu L_2}^{*}(\k)
(f_\mu ^{\dagger })^{
\bar{a}}\epsilon _\xi ^{b\bar{a}}\langle Z^\xi \rangle G_{L_2}^{c\eta
^{\dagger }}(\k,i\omega ),
\label{arf1:eq}
\end{eqnarray}
\begin{eqnarray}
&&\hspace{-0.5cm}
[ i\omega -\Delta _a]G_a^{X\eta ^{\dagger }}(\k,i\omega )=i\delta
(t-t^{\prime })\langle \{X_j^a,\eta _{j^{\prime }}^{\dagger }\}\rangle_\k
\nonumber \\
&&\hspace{-0.4cm}+\left[{\cal O}_{\mu L_1}^{-1}(\k)h_{L_1L_2}^c(\k)+
{\cal O}_{\mu \mu _1}^{-1} V_{\mu
_1L_2}^{*}(\k)\right]\nonumber \\ 
&&\hspace{-0.4cm}\times(f_\mu^{\dagger })^{\bar{b}}\epsilon _\xi ^{\bar{b}
a}\langle Z^\xi \rangle G_{L_2}^{c\eta ^{\dagger }}(\k,i\omega )  \nonumber \\
&&\hspace{-0.4cm}+{\cal O}_{\mu _1L_1}^{-1}(\k)V_{L_1\mu }(\k)
(f_{\mu _1}^{\dagger })^{\bar{b}
}\epsilon _\xi ^{\bar{b}a}\langle Z^\xi \rangle (f_\mu)^aG_a^{X\eta
^{\dagger }}(\k,i\omega ).
\label{arf2:eq}
\end{eqnarray}
Let us now try to identify if we, from the general equations, can 
establish some connection to ESC. 
In ESC calculations $n$ 
orbitals are localized and (for $f$-systems) $14-n$ orbitals are 
delocalized. This picture naturally arises in our approach within the 
orbital representation, where the wave function is taken in the simple form 
$|\Gamma _n\rangle =f_{\mu _1}^{\dagger }f_{\mu _2}^{\dagger }...f_{\mu
_n}^{\dagger }|0\rangle \equiv |\tilde{0})$. Since in this case the occupied 
orbitals are fixed, there are only one such state~\footnote{
Without coupling to the lattice this state may be degenerate.}. 
Then, 
the states with $(n+1)$ electrons arise when one of the orbitals $\varphi
_{\nu _1},\varphi _{\nu _2},...,\varphi _{\nu _{14-n}}$ , which are empty in 
the state $|\tilde{0}),$ is occupied in the state $|\Gamma _{n+1}\rangle
=f_\nu ^{\dagger }|\tilde{0})$. 
Any of the states $|\Gamma _{n+1}\rangle $ contains 
only one $f$-orbital of the type $\nu $ (which is delocalized due to its 
overlap and mixing with the conduction-electron orbitals), each particular 
transition $a:|\Gamma _{n+1}\rangle \rightarrow |\Gamma _n\rangle $, or, 
more explicitly, $f_\nu ^{\dagger }|\tilde{0})\rightarrow |\tilde{0})$, 
described by the Hubbard operator $X^a$, in this representation can be 
relabeled by $X^\nu$ since it fills the orbital $\nu$. Thus, one can indeed 
establish a one-to-one correspondence between filling unoccupied $f$-orbitals 
and the many-electron transitions. But this is possible only in this 
particular representation, which ignores the Hund rules and, therefore, 
ignores an essential part of the problem. 
Moreover, even in this representation the degeneracy must enter the equations
explicitly, since it always exist
if we ignore the intra shell spin-spin exchange (first Hund rule) 
and orbital-orbital exchange (second Hund rule). 
For the less than half filled shell 
this is not very essential unless we consider spin 
excitations and do not take into account the crystal field since the 
requirement of maximal spin and orbital moment makes the ground-state wave 
function unique. 
Let us rewrite our
equations
in the following
matrix form, taking into account the above assumption (here, for a brevity we 
omit the $\k$-dependence in $h(\k), V(\k)$, and ${\cal O}(\k)$):
\begin{eqnarray}
&&\left(
\begin{array}{cc}
\lbrack \delta _{LL_2}i\omega -{\cal O}_{L,L_1}^{-1}h_{L_1L_2}^c-{\cal O}_{L\nu
_1}^{-1}(f_{\nu_1})^aV_{\nu L_2}^{*}(f_\nu^{\dagger})^{\bar{a}}P^a] &
[{\cal O}_{L,\nu }^{-1}(f_\nu)^a\Delta _2+{\cal O}_{LL_1}^{-1}V_{L_1\nu }
(f_\nu)^a] \\
\lbrack {\cal O}_{\nu L_1}^{-1}h_{L_1L_2}^c+{\cal O}_{\nu \nu _1}^{-1}V_{\nu
_1L_2}^{*}](f_\nu ^{\dagger })^{\bar{a}}P^a & [i\omega -\Delta _a-{\cal O}_{\nu
_1L_1}^{-1}V_{L_1\nu }(f_{\nu _1}^{\dagger })^{\bar{a}}P^a(f_\nu)^a]
\end{array}
\right) \nonumber \\
&&\times\left(
\begin{array}{cc}
G_{L_2L^{\prime }}^{cc^{\dagger }}(i\omega ) & G_{L_2\bar{a}^{\prime
}}^{cX^{\dagger }}(i\omega ) \\
G_{aL^{\prime }}^{Xc^{\dagger }}(i\omega ) & G_{a\bar{a}^{\prime
}}^{XX^{\dagger }}(i\omega )
\end{array}
\right) =i\left(
\begin{array}{cc}
{\cal O}_{L,L^{\prime }}^{-1} & {\cal O}_{L\nu _1}^{-1}
(f_{\nu_1})^{a^{\prime }}P^a \\
{\cal O}_{\nu L^{\prime }}^{-1}(f_\nu ^{\dagger })^{\bar{a}}P^a &
{\cal O}_{\nu\nu^{\prime }}^{-1}P^a
\end{array}
\right) .
\end{eqnarray}
The on-site matrix elements $(f_\nu ^{\dagger })^{\bar{a}_{\nu_1}}$, which in 
the case of an orthogonal basis set give ''selection rules'', give a 
non-zero value: $(f_\nu ^{\dagger })^{\bar{a}_{\nu_1}}=
(\tilde{0}|f_{\nu_1}\cdot f_\nu ^{\dagger}|\tilde{0})=
{\cal O}_{\nu_1\nu}^{-1}$. 
For example, for the case 
of one $c$-orbital and two delocalized $f$-orbitals ${\cal O}_{\nu _1\nu
}^{-1}=[(1-\delta _{\nu \nu _1}){\cal O}_{\nu c}{\cal O}_{c\nu _1}+
\delta_{\nu \nu
_1}({\cal O}_{cc}-|{\cal O}_{\nu c}|^2)]/\det {\cal O}$, where 
$\det {\cal O}={\cal O}_{cc}-|{\cal O}_{\nu
c}|^2-|{\cal O}_{\nu _1c}|^2$. If we restrict ourselves with the accuracy 
$\sim o({\cal O}^1),$ i.e.\ neglect the matrix elements 
$\sim |{\cal O}_{\nu c}|^2$, then ${\cal O}_{\nu
_1\nu }^{-1}\simeq \delta _{\nu _1\nu }$. If we ignore the multiplet 
structure, then all energies $\Delta _a=\Delta _2$ coincide, and we find 
that the combination of population numbers $P^{[\Gamma _n,\Gamma
_{n+1}]}=S_{\Gamma _{n+1},\Gamma _{n+1}}N_{\Gamma _n}+S_{\Gamma _n,\Gamma
_n}N_{\Gamma _{n+1}}$  for different transitions, $a$, does not depend on 
$a$: $P^a=P$~\footnote{
In general, of course, this is a matrix $M\times M$, where $M$ is number of 
transitions.}. For the states $\Gamma _n$ we have $S_{\Gamma _n,\Gamma
_n}\equiv \langle \Gamma _n|\Gamma _n\rangle =(\tilde{0}|\tilde{0})=1$ and 
for $\Gamma _{n+1}$ $S_{\Gamma _{n+1},\Gamma _{n+1}}\equiv \langle \Gamma
_{n+1}|\Gamma _{n+1}\rangle =(\tilde{0}|f_\nu f_\nu ^{\dagger }|\tilde{0}
)\simeq 1.$ With these approximations we obtain the following system of 
equations 
\begin{eqnarray}
&&\left(
\begin{array}{cc}
\lbrack \delta _{LL_2}i\omega -{\cal O}_{L,L_1}^{-1}h_{L_1L_2}^c-{\cal O}_{L\nu
}^{-1}V_{\nu L_2}^{*}P] & [{\cal O}_{L,\nu }^{-1}\Delta _2+
{\cal O}_{LL_1}^{-1}V_{L_1\nu }]
\\
\lbrack {\cal O}_{\nu L_1}^{-1}h_{L_1L_2}^c+{\cal O}_{\nu \nu _1}^{-1}
V_{\nu _1L_2}^{*}]P
& [i\omega -\Delta _a-{\cal O}_{\nu _1L_1}^{-1}V_{L_1\nu }P]
\end{array}
\right)
\left(
\begin{array}{cc}
G_{L_2L^{\prime }}^{cc^{\dagger }}(\k,i\omega ) & G_{L_2\nu ^{\prime
}}^{cX^{\dagger }}(\k,i\omega ) \\
G_{\nu L^{\prime }}^{Xc^{\dagger }}(\k,i\omega ) & G_{\nu \nu ^{\prime
}}^{XX^{\dagger }}(\k,i\omega )
\end{array}
\right) \nonumber \\ 
&&=i\left(
\begin{array}{cc}
{\cal O}_{L,L^{\prime }}^{-1} & {\cal O}_{L\nu }^{-1}P \\
{\cal O}_{\nu L^{\prime }}^{-1}P & \delta _{\nu \nu ^{\prime }}P
\end{array}
\right) .
\label{finale:eq}
\end{eqnarray}
Let us now introduce the effective Fermions $\tilde{f}\equiv X/\sqrt{P}$ with
the help of the following transformation:
\begin{equation}
(\mathbf{I}\omega -\mathbf{\Omega })\mathbf{G}=\mathbf{P}\;\Rightarrow \;
\mathbf{U}(\mathbf{I}\omega -\mathbf{\Omega })\mathbf{U}^{-1}\mathbf{g}=
\mathbf{UPU,}
\end{equation}
where the block-matrix $\mathbf{U}$ is:
\begin{equation}
\mathbf{U=}\left(
\begin{array}{cc}
1 & 0 \\
0 & \frac 1{\sqrt{P}}
\end{array}
\right)
\end{equation}
and $\mathbf{g=UGU}$. After this transformation Eq.(\ref{finale:eq}) 
acquires the following form:
\begin{eqnarray}
\label{sce:eq}
&&\left(
\begin{array}{cc}
\lbrack \delta _{LL_2}i\omega -{\cal O}_{L,L_1}^{-1}h_{L_1L_2}^c-
({\cal O}_{L\nu
}^{-1}\sqrt{P})(V_{\nu L_2}^{*}\sqrt{P})] & [({\cal O}_{L,\nu }^{-1}
\sqrt{P})\Delta
_2+{\cal O}_{LL_1}^{-1}(V_{L_1\nu }\sqrt{P})] \\
\lbrack {\cal O}_{\nu L_1}^{-1}h_{L_1L_2}^c+{\cal O}_{\nu \nu _1}^{-1}
V_{\nu_1L_2}^{*}]
\sqrt{P}
& [i\omega -\Delta _a-({\cal O}_{\nu _1L_1}^{-1}\sqrt{P})(V_{L_1\nu }\sqrt{P})]
\end{array}
\right)
\left(
\begin{array}{cc}
g_{L_2L^{\prime }}^{cc^{\dagger }}(k,i\omega ) & g_{L_2\nu ^{\prime
}}^{cX^{\dagger }}(k,i\omega ) \\
g_{\nu L^{\prime }}^{Xc^{\dagger }}(k,i\omega ) & g_{\nu \nu ^{\prime
}}^{XX^{\dagger }}(k,i\omega )
\end{array}
\right)   \nonumber \\
&&=i\left(
\begin{array}{cc}
{\cal O}_{L,L^{\prime }}^{-1} & {\cal O}_{L\nu }^{-1}\sqrt{P} \\
{\cal O}_{\nu L^{\prime }}^{-1}\sqrt{P} & \delta _{\nu \nu ^{\prime }}
\end{array}
\right) .
\end{eqnarray}

Let us compare the system of equations, Eq.(\ref{sce:eq}), for the 
GFs, $G$, 
written in orbital
representation, with the system of equations for the Fermion GFs, $F$, 
corresponding to the secular problem in normal electronic structure
calculations, 
Eq.(\ref{bsc:eq}). 
We find that the systems coincide if we renormalize mixing and overlap
matrices as follows:
\begin{equation}
\label{rotenP:eq}
{\cal O}_{L\nu _1}^{-1}\rightarrow \tilde{{\cal O}}_{L\nu _1}^{-1}=
{\cal O}_{L\nu _1}^{-1}\sqrt{P}
,\;V_{L\nu _1}\rightarrow \tilde{V}_{L\nu _1}=V_{L\nu _1}\sqrt{P},
\end{equation}
and put 
${\cal O}_{\nu \nu ^{\prime }}^{-1}=\delta_{\nu\nu^{\prime }}$.
Thus, the matrix equation for the renormalized GFs, 
$(\mathbf{I}\omega -\mathbf{
\tilde{{\cal O}}}^{-1}\mathbf{\tilde{V}})\mathbf{g}=
\mathbf{\tilde{{\cal O}}}^{-1}$, can
be rewritten in the form containing the standard secular matrix used in the
electronic structure calculations
\begin{equation}
(\mathbf{\tilde{{\cal O}}}\omega -\mathbf{\tilde{V}})\mathbf{g}=\mathbf{I.}
\end{equation}
Now, using the standard procedure of Cholesky decomposition, $\mathbf{I=}(
\mathbf{\tilde{{\cal O}}}\omega -\mathbf{\tilde{V}})\mathbf{g}=
(\mathbf{Z}\omega
\mathbf{\bar{Z}}-\mathbf{\tilde{V}})\mathbf{g}$, we can rewrite this
equation in terms of an effective Hamiltonian:
\begin{equation}
\mathbf{(I\omega -Z}^{-1}\mathbf{\tilde{V}\bar{Z}}^{-1}\mathbf{)\bar{Z}gZ=I,}
\end{equation}
or,
\begin{equation}
\mathbf{(I\omega -h)\tilde{g}=I.}
\end{equation}
Thus, diagonalizing this equation in $\k$-space, we find the renormalized
band structure.
The parameter $P$, is calculated from the Green function $G^{XX}$.

In the paper II~\cite{igor_totale} we have shown that within the ARF 
each graph 
contributing to the Green function and the self-energy in the  
weak coupling perturbation theory has its counter-part in the 
strong-coupling theory and, therefore, one can use 
the same expression for the total energy in the ARF 
as in the case of weak-coupling 
theory but with the renormalizations shown in Eq.(\ref{rotenP:eq}).

\section{Electronic structure calculation in the ARF}
Since the secular equations to be solved include, in an explicit way, 
many body corrections in ARF, 
we will refer to this calculation as an 
electronic structure calculation in ARF (ESC-ARF). 
Let us now describe the numerical implementation of the ESC-ARF method. 
Following
the analysis of the previous section, as well as our previous 
work~\cite{lundin}, the
f-states are divided into a lower level and an upper level.
In ESC the electrons are normally separated into core and valence electrons, 
and it 
is not hard, in practice, to include the many body corrections derived in 
the previous section.
The lower 
transitions are simply single electron core levels in the ESC, and the upper 
transitions are all described by the set of equations described in the 
previous section.
Due to the analogy made between the obtained equations, Eq.(\ref{sce:eq})
and the standard secular equation solved for the Kohn-Sham
equation, Eq.(\ref{bsc:eq}), one needs to perform only relatively small 
modifications to any electronic structure method. 
Implemented in a full potential, site centered basis function method (to be 
specific, a 
linear muffin-tin orbital method) the 
Hamiltonian matrix has the schematic form as is shown in 
Fig.~\ref{hamil:fig}.
The energy of the upper transition, $\Delta_2$ may be used as a variable 
input parameter, 
but it can also 
be taken from bremsstrahlung isochromat spectra (BIS). The values obtained in 
this way are presented in Table~\ref{BIS:tab}.
\begin{table}[hbt]
\begin{tabular}{cccccc}
\hspace*{5mm}element & \hspace*{5mm}Ce & \hspace*{5mm}Pr &
\hspace*{5mm} Nd &\hspace*{5mm} Pm\tablenotemark[1] &\hspace*{5mm} Sm \\ \hline 
\hspace*{5mm}\f-peak & \hspace*{5mm}3.52 &\hspace*{5mm} 2.27 &
\hspace*{5mm} 1.68 &\hspace*{5mm} 1.08\tablenotemark[1] &\hspace*{5mm} 0.64 
\\
\end{tabular}
\tablenotetext[1]{Pm is radioactive, this value is a fit between Nd and Sm}
\caption{Positions of the \f-peak above the Fermi level in the lanthanide 
series in eV, taken from BIS, from ref~\cite{baer}}
\label{BIS:tab}
\end{table}
The renormalizing factor, $\sqrt{P}$, 
which should multiply the Hamiltonian and the overlap matrices in the places 
specified in Fig.~\ref{hamil:fig} has to be calculated self-consistently. 
Since we only 
take into account the states, $\Gamma_n$ and $\Gamma_{n+1}$, there are only 
two non-zero population numbers, $N_{\Gamma_n}$ and $N_{\Gamma_{n+1}}$, which 
gives rise to a single $P$. Further, we disregard the multiplet structure 
and therefore all transitions have the same energy, $\Delta_1$ for the 
lower level
and $\Delta_2$ from the upper level. The population numbers distribute as 
follows;
\begin{equation}
\label{eta:eq}
\left\{
\begin{array}{l}
n_f = n\cdot N_{1} + (n+1)\sum_{\nu} N_{2} \deff n+\eta  \\
N_{1} + \sum_{\nu} N_{2} \equiv 1, \nonumber 
\end{array}
\right.
\end{equation}
where $n$ and $n+1$ are the degeneracy of the states $\Gamma_n$ and 
$\Gamma_{n+1}$,
respectively,  and $\nu$ is an index of the available single-electron orbitals 
to construct 
$\Gamma_{n+1}$ from $\Gamma_n$.
If we assume that all upper transitions (this number is the same as the 
number of 
available single-electron orbitals) are occupied symmetrically, we find that 
$\sum_{\nu} N_{2}=nN_2$, 
i.e.\ $N_{2}=\eta/(14-n)$ and $N_{1} = 1-\eta$. Then, the renormalizing factor 
will be
\begin{equation}
P=1-\frac{14-(n+1)}{14-n}\eta.
\end{equation}
The \f-electron locator GF becomes 
\begin{equation}
{\mathcal G}_1 ={\mathcal G}_2 = \ldots = 
{\mathcal G}_n=\frac{1-\eta}{i\omega_n - \Delta_1} +
\frac{\eta}{i\omega_n - \Delta_2}
\end{equation}
and
\begin{equation}
{\mathcal G}_{\nu} = \frac{1-\frac{14-(n+1)}{14-n}\eta}{i\omega_n-\Delta_2} +
\frac{\frac{14-(n+1)}{14-n}\eta}{i\omega_n- \Delta_3}.
\label{gnu:eqn}
\end{equation}
Here, $\Delta_3$ is the energy of the $n+1$ to $n+2$ transition, which is much 
above the Fermi level and can be ignored. 
As seen the equations above contain an
unknown factor $\eta$. This factor should be 
calculated from the Green function
\begin{equation}
\eta=-\frac{1}{\pi}\sum_{\omega,\nu,{\bf k} }  Im \left (
{\mathcal G}_\nu
(\omega+i\delta,{\bf k})f(\omega)
\right ).
\end{equation}
This equation has to be solved self-consistently, since the 
factor $P$ renormalizes the Hamiltonian and overlap. 

The corrections has been implemented in a full potential linear muffin tin 
(FP-LMTO) method~\cite{wills,wills87}. For the lanthanides, the 
face centered cubic structure (fcc) cubic lattice was used. This was due to 
the simplicity to perform calculations with one atom per primitive cell.
In our calculations we set the same criteria for convergence in k-point 
summation,
self-consistency, basis set truncation etc. as is done in normal electronic 
structure calculations.

\section{Results}
We selected Ce, Pr, Nd and Sm as elements of special interest to test 
our theory for, since we, in a previous
work showed that the many-body 
effect of the upper transition is largest for them.
The renormalizing factor, $\sqrt{P}$, is, according to our calculations,
less than unity, and, thus, the 
mixing matrix elements are reduced somewhat. This means that the bandwidth is 
reduced compared to that obtained from a regular ESC. 
The upper transition can be taken from experiment, when available, or 
alternatively 
calculated from a constrained total energy calculation. In the present 
investigation
we also made a scan, changing 
$\Delta_2$, from a position high above the Fermi-level, to a place closer to 
the Fermi-level. The results of the ESC-ARF theory
are shown in Fig.~\ref{vol_f_P:fig}.
In this figure, the top figure shows the volumes as a function of the
position of the upper transition, $\Delta_2$. The central one, shows the 
f-occupation and the lower one shows the renormalization factor, $\sqrt{P}$, 
for Ce, Pr, Nd and Pm. We note that the renormalizing factor is quite 
close to unity for all cases which means that the effect of the correlations 
 for these materials is quite small. 
Further, when the upper transition is lowered, 
the \f-occupation increases, and there is a contribution to the cohesive energy 
from the \f-electrons, that results in that the volume decreases. 
In Table~\ref{vol:tab} we list the volumes for the selected lanthanides. 
\begin{table}[hbt]
\begin{tabular}{cccc}
\hspace*{5mm}element  & \hspace*{5mm}Expt.\tablenotemark[1] \hspace*{5mm}& 
 LDA\tablenotemark[2] \hspace*{5mm}& here \hspace*{5mm}\\ \hline
\hspace*{5mm}Ce       & \hspace*{5mm}3.8115 \hspace*{5mm}& 
3.8994 \hspace*{5mm}& 3.7485 \hspace*{5mm}\\
\hspace*{5mm}Pr       & \hspace*{5mm}3.8178 \hspace*{5mm}& 3.8375 
\hspace*{5mm}& 3.9030 \hspace*{5mm}\\
\hspace*{5mm}Nd       & \hspace*{5mm}3.8045 \hspace*{5mm}& 3.7848 
\hspace*{5mm}& 3.7484 \hspace*{5mm}\\
\hspace*{5mm}Pm\tablenotemark[3] & \hspace*{5mm}3.7832 \hspace*{5mm}& 
3.7415 \hspace*{5mm}& 3.5182 \hspace*{5mm}\\
\end{tabular}
\tablenotetext[1]{Taken from ref.~\cite{gschneider90}.}
\tablenotetext[2]{ESC with the 4\f~treated as core electrons.}
\tablenotetext[3]{Pm is radioactive, this value is a fit between Nd and Sm.}
\caption{Volumes, R$_{{\rm SWS}}$, for the light lanthanides, in a.u. When 
the position of the upper transition, $\Delta_2$ is put to the experimental 
value shown in Table~\ref{BIS:tab}}
\label{vol:tab}
\end{table}

In Fig.~\ref{dos:fig}, the density of states after the self-consistent cycle 
is shown. The occupied part of the $d$-band is quite similar for the 
case of a normal calculation treating the f-level as part of the core. 
Also, the
4\f-band like feature is rather similar to what would be obtained from 
a regular ESC
calculation, with a small narrowing due to the renormalization and a 
shift to higher energies,
so that the number of f-electrons occupying this feature is small, 
$\sim$ 0.05-0.1. The 
resulting electronic structure of our ESC-ARF 
theory is actually rather similar to that
obtained when including Hubbard-Kanamory interactions to an LDA 
Hamiltonian (or the related LDA+U
method) and to what is obtained from calculations including 
self-interaction corrections.

In Fig.~\ref{f-width:fig} the f-bandwidth as a function of the position of the 
upper transition with respect to the Fermi level is plotted. The circles 
correspond to a self-consistent calculation where $\sqrt{P}$ is put 
to 1 during the 
whole cycle. The boxes are the result from a full calculation including all 
renormalization effects described above. 
A band narrowing effect is seen when $\sqrt{P}\neq 1$, 
but one may note that the effect is
not extremely pronounced for the presently studied elements. 
One may speculate that
for other systems, where a larger population of the upper level is expected 
(say, the level $\Delta _2$ is much closer to Fermi energy), 
the renormalization is
more pronounced. Since the present modifications are similar 
to those presented for
$\delta$-Pu\cite{eriksson}, that does show a larger 
population of the upper band, it may very well be so
that the present corrections are most important for this element. 

\section{conclusion}
We have presented a method to incorporate strong correlations into electronic 
structure calculations (ESC-ARF) and performed numerical tests of the theory. 
The results show a narrowing of the 4\f-bands compared to LDA (or GGA) 
results. Due to the 
division of the 4\f-states in a higher and lower level, we find that the 
cohesion from 
the 4\f-states is non-zero (in contrast to calculations that treat 
the 4\f-states 
as core states) but substantially smaller than what it would have been in 
an LDA (or GGA) calculation. 

The present results are rather similar to our previous calculations, that 
include modifications in 
an LMTO-ASA method, although the full-potential treatment does include 
finer details of the potential. 
Due to the generalization of the corrections to an electronic 
structure method, 
that can treat any crystal structure, we believe the present method 
should have a potential in
calculating the electronic structure of \f-based materials that are hard to 
treat in a regular LDA (or GGA) ESC methods.

\section{Acknowledgments}
We are thankful for financial support from
the Swedish Natural Science Research Council (NFR and TFR), for the 
foundation for
strategic research (SSF) and the G\"{o}ran Gustafsson foundation. 
J.M.\ Wills are acknowledged for supplying the codes for the FP-LMTO 
calculations.

\begin{figure}[hbt]
\caption{The Hamiltonian (and overlap)  matrix modified according to the
text. $\Delta_2$ is the position of the upper transition.}
\label{hamil:fig}
\end{figure}

\begin{figure}[hbt]
\caption{The volumes, f-occupation and $\sqrt{P}$-factor for
Ce, Pr, Nd and Pm, as a function of the position of the upper
transition,
$\Delta_2$.}
\label{vol_f_P:fig}
\end{figure}

\begin{figure}[hb]
\caption{The $d$- and \f-projected density of states for the two cases
when
there are f-states in the valence basis with all corrections described
above,
and when a normal ESC with the \f-electrons treated as core.}
\label{dos:fig}
\end{figure}

\begin{figure}[hb]
\caption{The bandwidth of the \f-states, when the renormalizing factor
is put
to one and when it is not. }
\label{f-width:fig}
\end{figure}

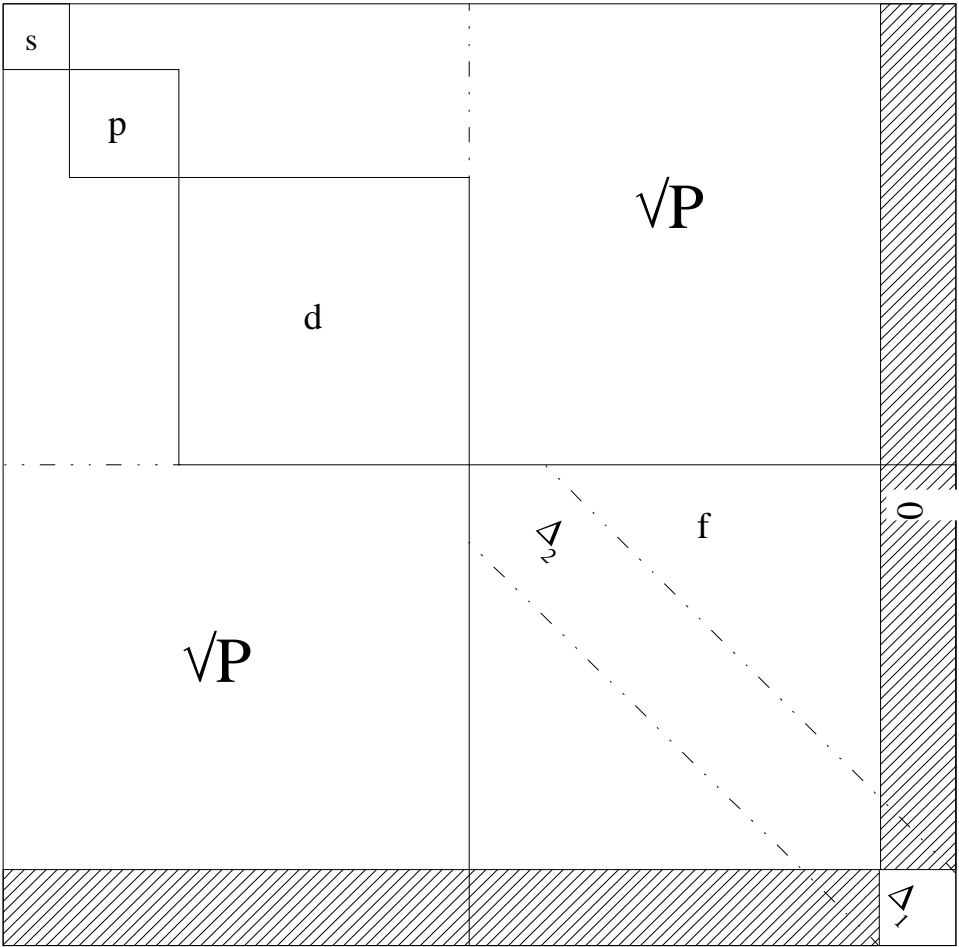
\end{document}